\documentclass[11pt]{article}
\usepackage[russian]{babel}
\usepackage{amssymb,amsmath,amsthm}
\usepackage{graphicx}
\usepackage{epstopdf}
\usepackage{setspace}
\newcommand{\bc}{\begin{center}}
\newcommand{\ec}{\end{center}}
\newcommand{\be}{\begin{equation}}
\newcommand{\ee}{\end{equation}}
\newcommand{\ba}{\begin{array}}
\newcommand{\ea}{\end{array}}
\newcommand{\edc}{\end{document}}

\begin{document}
\leftline{УДК 517.98}
\begin{center}
\textbf{\Large {Условия единственности и неединственности слабо периодических мер Гиббса для НС-модели}}
\end{center}

\begin{center}
Р.М.Хакимов\footnote{Наманганский государственный университет, Наманган, Узбекистан.\\
E-mail: rustam-7102@rambler.ru},
М.Т.Махаммадалиев\footnote{Наманганский государственный университет, Наманган, Узбекистан.\\
E-mail: mmtmuxtor93@mail.ru}
\end{center}

\begin{abstract}
В данной статье изучается Hard-Core (НС) модель на дереве Кэли. В случае нормального делителя индекса четыре при некоторых условиях на параметры показана единственность слабо периодических мер Гиббса. Кроме того, доказано существование слабо периодических (не периодических) мер Гиббса, отличных от ранее известных.
\end{abstract}

\textbf{Ключевые слова}: дерево Кэли, конфигурация, НС-модель,
мера Гиббса, трансляционно-инвариантная мера, периодическая мера, слабо периодическая мера.\

\section{Введение}\

Решения проблем, возникающих в результате исследований при изучении
термодинамических свойств физических и биологических систем, в основном
приводятся к задачам теории мер Гиббса. Мера Гиббса-это фундаментальный закон, определяющий
вероятность микроскопического состояния данной физической системы. Известно, что каждой
мере Гиббса сопоставляется одна фаза физической системы, и если мера Гиббса не единственна,
то говорят, что существует фазовый переход. Для достаточно широкого класса гамильтонианов известно, что множество всех предельных мер Гиббса (соответствующих данному гамильтониану) образует непустое выпуклое компактное подмножество в множестве всех вероятностных мер (см. например [1]-[3]), и каждая точка этого выпуклого множества однозначно разлагается по его крайним точкам. В связи с этим особый интерес представляет описание всех крайних точек этого выпуклого множества, т. е. крайних мер Гиббса.

Определение меры Гиббса и других понятий, связанных с теорией мер Гиббса, можно
найти, например, в работах [1]-[4]. Несмотря на многочисленные работы, посвященных изучению мер Гиббса, ни для одной модели не было получено полное описание всех предельных мер Гиббса.
Относительно других моделей для модели Изинга на дереве Кэли эта задача изучена достаточно полностью. Так,
например, в работе [5] построено несчетное множество крайних
гиббсовских мер, а в работе [6] найдены необходимое и
достаточное условия крайности неупорядоченной фазы модели Изинга
на дереве Кэли.

В работе [7] Мазелью и Суховым была введена и изучена НС-модель (жесткий диск, жесткая сердцевина) на $d$-мерной решетке $ \mathbb Z ^ d$.
Работы [8]-[15] посвящены к изучению мер Гиббса для HC-модели с двумя состояниями на дереве Кэли. В работе [8] была доказана единственность трансляционно-инвариантной меры и неединственность
периодических мер Гиббса для НС-модели. Также в [8]
(соответственно в [9]) найдено достаточное условие на
параметры НС-модели, при котором трансляционно-инвариантная мера Гиббса является некрайней
(соответственно крайней). В работе [10] расширена область крайности этой меры. Работа [11] посвящена изучению слабо периодических мер Гиббса для HC-модели для нормального делителя индекса два и при некоторых условиях на параметры показана единственность слабо периодической меры Гиббса, а в работе [12] дано полное описание слабо периодических мер Гиббса для HC-модели при любых значениях параметров в случае нормального делителя индекса два. В [13] доказано существование слабо периодических (не периодических) мер Гиббса для HC-модели для нормального делителя индекса четыре при некоторых условиях на параметры. В [14] для HC-модели с двумя состояниями улучшен один из результатов из работы [13]. В работе [15] для HC-модели в случае нормального делителя индекса четыре  доказано существование слабо периодических (не периодических) мер Гиббса, отличных от ранее известных.

Работы [16]-[19] посвящены изучению гиббсовских мер для HC-моделей с тремя состояниями на
дереве Кэли порядка $k\geq1$. Для ознакомления с другими свойствами НС-модели (и их обобщениями)
на дереве Кэли см. Главу 7 монографии [4].

В данной работе изучается НС-модель с двумя состояниями на дереве Кэли. В случае нормального делителя индекса четыре при некоторых условиях на параметры показана единственность слабо периодических мер Гиббса. Кроме того, улучшен один из результатов из работы [13]: найдено точное критическое значение параметра $\lambda_{cr}$ такое, что на дереве Кэли порядка три при $\lambda>\lambda_{cr}$ существуют ровно пять слабо периодические меры Гибсса, одна из которых является трансляционно-инвариантной, а четыре другие слабо периодическими (не периодическими). А также при некоторых условиях найдены явные значения $\lambda_{cr}^{(1)}(k)$ и $\lambda_{cr}^{(2)}(k)$, что при $\lambda_{cr}^{(1)}<\lambda<\lambda_{cr}^{(2)}$ существуют не менее двух слабо периодических (не периодических) мер Гиббса на дереве Кэли порядка $k\geq6$.

\section{Предварительные сведения}\

Пусть $\Gamma^k=(V,L,i)$ есть дерево Кэли порядка $ k\geq 1$, где $V$ есть множество
вершин $\Gamma^k$, $L-$множество его ребер и $i-$функция
инцидентности, сопоставляющая каждому ребру $l\in L$ его концевые
точки $x, y \in V$. Если $i (l) = \{ x, y \} $, то $x$ и $y$
называются  {\it ближайшими соседями вершины} и обозначается $l =
\langle x,y\rangle $. Пусть $d(x,y), x, y \in V$ есть расстояние между вершинами $x, y$, т.е. количество ребер кратчайшей пути, соединяющей $x$ и $y$.

Для фиксированного $x^0\in V$ обозначим
$$ W_n =\{x\in V | \ d (x, x^0) =n \}, \  V_n = \cup_{j=0}^{n}W_j.$$
Для $x\in W_{n}$ обозначим (множество прямых потомков вершины $x$)
$$ S(x)=\{y\in{W_{n+1}}:d(x,y)=1\}.$$

Пусть $\Phi=\{0,1\}$ и $\sigma\in\Phi^V-$ конфигурация, то есть
$\sigma=\{\sigma(x)\in \Phi: x\in V\}$, где $\sigma (x)=1$
означает, что вершина $x$ на дереве Кэли занята, а $\sigma (x)=0$
означает, что она свободна. Конфигурация $\sigma$ называется
допустимой, если $\sigma (x)\sigma (y)=0$ для любых соседних
$\langle x,y \rangle $ из $V$ ($V_n $ или $W_n$, соответственно) и
обозначим множество таких конфигураций через $\Omega$
($\Omega_{V_n}$ и $\Omega_{W_n}).$ Ясно, что
$\Omega\subset\Phi^V.$

Объединение конфигураций $\sigma_{n-1}\in\Phi ^ {V_{n-1}}$ и $\omega_n\in\Phi
^ {W_{n}}$ определяется следующей формулой (см. [20])
$$
\sigma_{n-1}\vee\omega_n=\{\{\sigma_{n-1}(x), x\in V_{n-1}\},
\{\omega_n(y), y\in W_n\}\}.
$$

Гамильтониан HC-модели определяется по формуле
  $$H(\sigma)=\left\{%
\begin{array}{ll}
    J \sum\limits_{x\in{V}}{\sigma(x),} \ \ \ $ если $ \sigma \in\Omega $,$ \\
   +\infty ,\ \ \ \ \ \ \ \ \ \ $  \ если $ \sigma \ \notin \Omega $,$ \\
\end{array}%
\right. $$ где $J\in R$.

Пусть $\mathbf{B}$ есть $\sigma$-алгебра, порожденная
цилиндрическими подмножествами $\Omega.$ Для любого $n$ обозначим
через $\mathbf{B}_{V_n}=\{\sigma\in\Omega:
\sigma|_{V_n}=\sigma_n\}$ подалгебру $\mathbf{B},$ где
$\sigma|_{V_n}-$ сужение $\sigma$ на $V_n,$ $\sigma_n: x\in V_n
\mapsto \sigma_n(x)-$ допустимая конфигурация в $V_n.$

\textbf{Определение 1}. Для $\lambda >0$ НС-мера Гиббса есть
вероятностная мера $\mu$ на $(\Omega , \textbf{B})$ такая, что для
любого $n$ и $\sigma_n\in \Omega_{V_n}$
$$
\mu \{\sigma \in \Omega:\sigma|_{V_n}=\sigma_n\}=
\int_{\Omega}\mu(d\omega)P_n(\sigma_n|\omega_{W_{n+1}}),
$$
где
$$
P_n(\sigma_n|\omega_{W_{n+1}})=\frac{e^{-H(\sigma_n)}}{Z_{n}
(\lambda ; \omega |_{W_{n+1}})}\textbf{1}(\sigma_n \vee \omega
|_{W_{n+1}}\in\Omega_{V_{n+1}}).
$$

Здесь $Z_n(\lambda ; \omega|_{W_{n+1}})-$ нормировочный множитель с
граничным условием $\omega|_{W_n}$:
$$
Z_n (\lambda ; \omega|_{W_{n+1}})=\sum_{\widetilde{\sigma}_n \in
\Omega_{V_n}}
e^{-H(\widetilde{\sigma}_n)}\textbf{1}(\widetilde{\sigma}_n\vee
\omega|_{W_{n+1}}\in \Omega_{V_{n+1}}).
$$

Для $\sigma_n\in\Omega_{V_n}$ положим
$$\#\sigma_n=\sum\limits_{x\in V_n}{\mathbf 1}(\sigma_n(x)\geq 1)$$
число занятых вершин в $\sigma_n$.

Пусть $z:x\mapsto z_x=(z_{0,x}, z_{1,x}) \in R^2_+$
векторнозначная функция на $V$. Для $n=1,2,\ldots$ и $\lambda>0$
рассмотрим вероятностную меру $\mu^{(n)}$ на $\Omega_{V_n}$,
определяемую как
$$
\mu^{(n)}(\sigma_n)=\frac{1}{Z_n}\lambda^{\#\sigma_n} \prod_{x\in
W_n}z_{\sigma(x),x}.
\eqno(1)$$
Здесь $Z_n-$ нормирующий делитель:
$$Z_n=\sum_{{\widetilde\sigma}_n\in\Omega_{V_n}}
\lambda^{\#{\widetilde\sigma}_n}\prod_{x\in W_n}
z_{{\widetilde\sigma}(x),x}.$$

Говорят, что последовательность вероятностных мер $\mu^{(n)}$ является
согласованной, если для любых $n\geq 1$ и
$\sigma_{n-1}\in\Omega_{V_{n-1}}$:
$$
\sum_{\omega_n\in\Omega_{W_n}}
\mu^{(n)}(\sigma_{n-1}\vee\omega_n){\mathbf 1}(
\sigma_{n-1}\vee\omega_n\in\Omega_{V_n})=
\mu^{(n-1)}(\sigma_{n-1}).
\eqno(2)$$

В этом случае существует единственная мера $\mu$ на $(\Omega,
\textbf{B})$ такая, что для всех $n$ и $\sigma_n\in \Omega_{V_n}$
$$\mu(\{\sigma|_{V_n}=\sigma_n\})=\mu^{(n)}(\sigma_n).$$

\textbf{Определение 2.} Мера $\mu$, являющейся пределом последовательности $\mu^{(n)}$, определенной формулой
(1) с условием согласованности (2), называется HC-\textit{мерой Гиббса} с $\lambda>0$,
\textit{соответ-ствующей функции} $z:\,x\in V \setminus\{x^0\}\mapsto z_x$. При этом
HC-мера Гиббса, соответствующая постоянной функции $z_x\equiv z$,
называется трансляционно-инвариантной.\

Известно, что существует взаимнооднозначное соответствие между
множеством $V$ вершин дерева Кэли порядка $k\geq 1$ и группой
$G_k$, являющейся свободным произведением $k+1$ циклических групп
второго порядка с образующими $a_1,...,a_{k+1}$, соответственно
(см. [21]). Поэтому множество $V$ можно отождествлять
c множеством $G_k$.

Пусть $\widehat{G}_k-$ подгруппа группы $G_k$.

Если гиббсовская мера инвариантна относительно некоторой подгруппы
конечного индекса $ \widehat{G}_k\subset {G_k}$, то она называется
$\widehat{G}_k$-периодической.

Известно [8], что каждой мере Гиббса для
HC-модели на дереве Кэли можно сопоставлять совокупность величин $z=\{z_x, x\in
G_k \},$ удовлетворяющих
$$
z_x=\prod_{y \in S(x)}(1+\lambda z_y)^{-1},
\eqno(3)$$
где $\lambda=e^{J_1}>0-$
параметр, $J_1=-J\beta$, $\beta={1\over T}$, $T>0-$температура.\

\textbf{Определение 3}. Совокупность величин $z=\{z_x,x\in G_k\}$
называется $ \widehat{G}_k$-периодической, если  $z_{yx}=z_x$ для
$\forall x\in G_k, y\in\widehat{G}_k.$\

$G_k$-периодические совокупности называются
трансляционно-инвариантными.

Для любого $x\in G_k $ множество $\{y\in G_k: \langle
x,y\rangle\}\setminus S(x)$ имеет единственный элемент, которого
обозначим через $x_{\downarrow}$ (см.[22]).

Пусть $G_k/\widehat{G}_k=\{H_1,...,H_r\}$ фактор группа, где
$\widehat{G}_k-$нормальный делитель индекса $r\geq 1.$

\textbf{Определение 4}. Совокупность величин $z=\{z_x,x\in G_k\}$
называется $\widehat{G}_k$-слабо периодической, если
$z_x=z_{ij}$ при $x\in H_i, x_{\downarrow}\in H_j$ для $\forall
x\in G_k$.\

\textbf{Определение 5}. Мера $\mu$ называется
$\widehat{G}_k$-(слабо) периодической, если она соответствует
$\widehat{G}_k$-(слабо) периодической совокупности величин $z$.\

\section{Известные факты}\

Пусть $A\subset\{1,2,...,k+1\}$ и $H_A=\{x\in
G_k:\sum\limits_{i\in A}w_x(a_i)-$ четное число$ \}$, где
$w_x(a_i)-$ число буквы $a_i$ в слове $x\in G_k$,
${G_k}^{(2)}=\{x\in G_k: \mid x\mid-\mbox{четное число}\},$ где
$\mid x\mid-$ длина слова $x\in G_k$ и
${G_k}^{(4)}=H_A\cap{G_k}^{(2)}-$ нормальный делитель индекса 4.\

Рассмотрим фактор-группу $G_k/{G_k}^{(4)}=\{H_0, H_1, H_2, H_3\},$
где
$$H_0=\{x\in G_k: \sum\limits_{i\in
A}w_x(a_i)-\mbox{четно}, |x|-\mbox{четно}\}$$
$$H_1=\{x\in G_k: \sum\limits_{i\in
A}w_x(a_i)-\mbox{нечетно}, |x|-\mbox{четно}\}$$
$$H_2=\{x\in G_k: \sum\limits_{i\in
A}w_x(a_i)-\mbox{четно}, |x|-\mbox{нечетно}\}$$
$$H_3=\{x\in G_k: \sum\limits_{i\in
A}w_x(a_i)-\mbox{нечетно}, |x|-\mbox{нечетно}\}$$

Тогда в силу (3), после некоторых преобразований, получим
$$
\left\{%
\begin{array}{ll}
    z_{1}=\frac{(1+\lambda z_{7})^k}{((1+\lambda z_7)^{k/i}+\lambda z_8^{1-1/i})^i}\cdot\frac{1}{(1+\lambda z_2)^{k-i}} \\
    \\
    z_{2}=\frac{(1+\lambda z_8)^k}{((1+\lambda z_8)^{k/i}+\lambda z_7^{1-1/i})^i}\cdot\frac{1}{(1+\lambda z_1)^{k-i}} \\
    \\
    z_{7}=\frac{(1+\lambda z_1)^k}{((1+\lambda z_1)^{k/i}+\lambda z_2^{1-1/i})^i}\cdot\frac{1}{(1+\lambda z_8)^{k-i}} \\
    \\
    z_{8}=\frac{(1+\lambda z_2)^k}{((1+\lambda z_2)^{k/i}+\lambda z_1^{1-1/i})^i}\cdot\frac{1}{(1+\lambda z_7)^{k-i}} \\
\end{array}%
\right.
\eqno(4)$$
Здесь $i=|A|-$ мощность множества $A$.

Рассмотрим отображение $W:R^4 \rightarrow R^4,$ определенное
следующим образом:

$$
\left\{%
\begin{array}{ll}
    z_{1}^{'}=\frac{(1+\lambda z_{7})^k}{((1+\lambda z_7)^{k/i}+\lambda z_8^{1-1/i})^i}\cdot\frac{1}{(1+\lambda z_2)^{k-i}}
    \\[3mm]
    z_{2}^{'}=\frac{(1+\lambda z_8)^k}{((1+\lambda z_8)^{k/i}+\lambda z_7^{1-1/i})^i}\cdot\frac{1}{(1+\lambda z_1)^{k-i}}
    \\[3mm]
    z_{7}^{'}=\frac{(1+\lambda z_1)^k}{((1+\lambda z_1)^{k/i}+\lambda z_2^{1-1/i})^i}\cdot\frac{1}{(1+\lambda z_8)^{k-i}}
    \\[3mm]
    z_{8}^{'}=\frac{(1+\lambda z_2)^k}{((1+\lambda z_2)^{k/i}+\lambda z_1^{1-1/i})^i}\cdot\frac{1}{(1+\lambda z_7)^{k-i}} \\
\end{array}%
\right.
$$

Заметим, что (4) есть уравнение $z=W(z).$ Чтобы решить систему уравнений (4), надо
найти неподвижные точки отображения $z^{'}=W(z).$\

Из [13] и [15] известны следующие леммы 1,2 и теоремы 1,2.

\textbf{Лемма 1.} [13] \textit{ Отображение $W$ имеет инвариантные
множества следующих видов:}
$$I_1=\{(z_1, z_2, z_7, z_8) \in R^4: z_1=z_2=z_7=z_8\}, \ \ I_2=\{(z_1, z_2, z_7, z_8)\in R^4: z_1=z_7, \ z_2=z_8\},$$
$$I_3=\{(z_1, z_2, z_7, z_8) \in R^4: z_1=z_2, z_7=z_8\}, \ \ I_4=\{(z_1, z_2, z_7, z_8)\in R^4: z_1=z_8, \ z_2=z_7\}.$$

\textbf{Лемма 2.} [13] \textit{Если на инвариантных множествах
$I_2, I_3, I_4$ существуют слабо периодические меры Гиббса, то они
являются либо трансляционно-инвариантными, либо слабо
периодическими (не периодическими).}\

\textbf{Замечание 1.} Когда говорим, что на инвариантном множестве
$I_m$ существуют слабо периодические меры Гиббса, то здесь имеется ввиду существование слабо периодических мер Гиббса, соответствующих совокупности величин из инвариантного множества $I_m$.

\textbf{Теорема 1.} [13] \textit{Для HC-модели в случае нормального
делителя индекса четыре верны следующие утверждения: }

\textit{1. При $k\geq1, i\leq k$ на $I_1$ слабо периодическая мера
Гиббса единственна. Более того, эта мера совпадает с единственной
трансляционно-инвариантной мерой Гиббса.}

\textit{2. Пусть $k=2, i=1, \lambda_{cr}=4$. Тогда на $I_2$ при
$\lambda<\lambda_{cr}$ существует одна слабо периодическая мера
Гиббса, которая является трансляционно-инвариантной, при
$\lambda=\lambda_{cr}$ существуют две слабо периодические меры
Гибсса, одна из которых является трансляционно-инвариантной,
другая слабо периодической (не периодической) и при
$\lambda>\lambda_{cr}$ существуют не менее двух слабо
периодических (не периодических) мер Гибсса.}

\textit{3. Пусть $k=3, i=1$. Тогда существует $\lambda_0$ такая,
что на $I_2$ при $\lambda>\lambda_{0}$ существуют не менее четырех
мер Гиббса, одна из которых является трансляционно-инвариантной, а
остальные слабо периодическими (не периодическими) мерами Гиббса.}

\textit{4. При $k\geq1, i=1$ на $I_3$ слабо периодическая мера
Гиббса единственна.}

\textit{5. При $k=2,3, i=1$ на $I_4$ слабо периодическая мера
Гиббса единственна.}\

\textbf{Замечание 2.} В работе [14] улучшен 2-пункт Теоремы 1, т.е. доказано, что при
$\lambda>\lambda_{cr}$ существуют ровно две слабо периодические (не периодические) меры Гибсса.\

\textbf{Теорема 2.} [15] \textit{Для HC-модели в случае нормального
делителя индекса четыре верны следующие утверждения: }

\textit{1. Пусть $k=2, i=2, \lambda_{cr}=4$. Тогда на $I_2$ при
$\lambda<\lambda_{cr}$ существует одна слабо периодическая мера
Гиббса, которая является трансляционно-инвариантной, при
$\lambda=\lambda_{cr}$ существуют две слабо периодические меры
Гибсса, одна из которых является трансляционно-инвариантной,
другая слабо периодической (не периодической) и при
$\lambda>\lambda_{cr}$ существуют ровно три слабо
периодические меры Гибсса, одна из которых является трансляционно-инвариантной,
а две другие слабо периодическими (не периодическими).}

\textit{2. При $k=2, i=2$ на $I_3$ слабо периодическая мера
Гиббса единственна. Более того, эта мера совпадает с единственной
трансляционно-инвариантной мерой Гиббса}

\textit{3. При $k=i$ на $I_4$ слабо периодическая мера
Гиббса единственна. Более того, эта мера совпадает с единственной
трансляционно-инвариантной мерой Гиббса.}\

\section{Условия единственности слабо периодических мер}\

Справедлива следующая теорема.

\textbf{Теорема 3.} \textit{Для HC-модели в случае нормального
делителя индекса четыре верны следующие утверждения: }

\textit{1. При $k=i$ слабо периодическая мера
Гиббса, соответствующая совокупности величин из множества $I_3$, единственна. Более того, эта мера совпадает с единственной трансляционно-инвариантной мерой Гиббса.}

\textit{2. В случаях}
$$1) \ k=3, \ i=2; \ 2) \ k=4, \ i=1; \  3) \ k=4, \ i=2; \ 4) \ k=4, \ i=3; \ 5) \ k=5, \ i=1$$
\textit{слабо периодическая мера Гиббса, соответствующая совокупности величин из множества $I_4$, единственна. Более того, эта мера совпадает с единственной трансляционно-инвариантной мерой
Гиббса.}\

\textbf{Доказательство.} \textbf{Случай $I_3$.} Пусть $k=i$. Рассмотрим  систему уравнений (4) на $I_3$:
$$
\left\{%
\begin{array}{ll}
    z_1=\left(\frac{1+\lambda z_7}{1+\lambda z_7+\lambda
z^{1-1/i}_7}\right)^i \\[3mm]
    z_7=\left(\frac{1+\lambda z_1}{1+\lambda z_1+\lambda
z^{1-1/i}_1}\right)^i. \\
    \end{array}%
\right.
\eqno(5)$$
Ясно, что $0<z_1<1$ и $0<z_7<1$. Введем обозначения:
$z_1=x^i$, $z_7=y^i$. Тогда система уравнений (5) имеет вид:
$$
\left\{%
\begin{array}{ll}
   x=\frac{1+\lambda y^i}{1+\lambda y^i+\lambda y^{i-1}}\\[3mm]
    y=\frac{1+\lambda x^i}{1+\lambda x^i+\lambda x^{i-1}} \\
    \end{array}%
\right.
\eqno(6)$$
или
$$
\left\{%
\begin{array}{ll}
  x+\lambda xy^i+\lambda xy^{i-1}=1+\lambda y^i\\[3mm]
    y+\lambda x^iy+\lambda x^{i-1}y=1+\lambda x^i,\\
    \end{array}%
\right.
\eqno(7)$$
где $0<x<1$ и $0<y<1$. Ясно, что если $y=x$, то мы получим решение, которое находится в $I_1$ и
оно единственно. Покажем, что система уравнений (7) (значит и система уравнений (6)) других решений не имеет.

Вычтем из первого уравнения (7) второе и обе части полученного уравнения разделим на $(y-x)$. В результате получим уравнение
$$-1+\lambda xy(y^{i-2}+xy^{i-3}+x^2y^{i-4}+...+x^{i-3}y+x^{i-2})+\lambda
xy(y^{i-3}+xy^{i-4}+x^2y^{i-5}+...+x^{i-4}y+x^{i-3})=$$
$$
=\lambda(y^{i-1}+xy^{i-2}+x^2y^{i-3}+...+x^{i-2}y+x^{i-1}).
\eqno(8)$$
Так как в уравнении (8) $y\neq x$, то $x>y$ или $y>x$. Предположим,
что $x<y$. Достаточно доказать, что при этом условии верно следующее неравенство:
$$\lambda xy(y^{i-2}+xy^{i-3}+x^2y^{i-4}+...+x^{i-3}y+x^{i-2})+\lambda
xy(y^{i-3}+xy^{i-4}+x^2y^{i-5}+...+x^{i-4}y+x^{i-3})<$$
$$
<\lambda(y^{i-1}+xy^{i-2}+x^2y^{i-3}+...+x^{i-2}y+x^{i-1})
\eqno(9)$$
Для этого воспользуемся методом матаматической индукции по $i$.

Пусть $i=3$. Тогда (9) после некоторых преобразований имеет вид
$$\lambda x^2y+\lambda xy^2<\lambda x^2+ \lambda y^2.$$
Заметим, что последнее неравенство верно, т.к. $0<x<1$ и $0<y<1$.

Далее, предположим, что неравенство (9) верно при $i=n-1$, т.е.
$$xy(y^{n-3}+xy^{n-4}+x^2y^{n-5}+...+x^{n-4}y+x^{n-3})+\lambda
xy(y^{n-4}+xy^{n-5}+x^2y^{n-6}+...+x^{n-5}y+x^{n-4})<$$
$$
<\lambda(y^{n-2}+xy^{n-3}+x^2y^{n-4}+...+x^{n-3}y+x^{n-2}).
\eqno(10)$$
Покажем, что это неравенство верно и при $i=n$.

Действительно, пусть $i=n$. Тогда неравенство (9) имеет вид
$$\lambda xy(y^{n-2}+xy^{n-3}+x^2y^{n-4}+...+x^{n-3}y+x^{n-2})+\lambda
xy(y^{n-3}+xy^{n-4}+x^2y^{n-5}+...+x^{n-4}y+x^{n-3})<$$
$$
<\lambda(y^{n-1}+xy^{n-2}+x^2y^{n-3}+...+x^{n-2}y+x^{n-1}).
\eqno(11)$$
Заметим, что следующее неравенство верно:
$$
-\lambda xy(x^{n-2}+x^{n-3})<-\lambda x^{n-1}.
\eqno(12)$$
Действительно, из неравенства (12) легко получить неравенство $x<y(x+1)$, а это верно, т.к. $x<y$, $0<x<1$ и $0<y<1$.

Далее, сложив правые и левые части неравенств (11) и (12), соответственно, и разделив обе части полученного неравенства на $y$, получим неравенство (10), которое справедливо по предположению.

В случае $x>y$ доказывается аналогично. Значит, при любых значений $\lambda>0$ система уравнений (6) имеет единственное решение вида $(x,x)$, которое соответствует единственной трансляционно-инвариантной мере Гиббса для НС модели, т.е. при $k=i$ слабо периодическая мера Гиббса, соответствующая совокупности величин из $I_3$, единственна и совпадает с единственной трансляционно-инвариантной мерой Гиббса. Первое утверждение теоремы доказано.\

\textbf{Случай $I_4$: $k=3$ и $i=2$.} Рассмотрим систему уравнений (4) на $I_4$.

При $k=3$ и $i=2$ она имеет вид:
$$
\left\{%
\begin{array}{ll}
  z_1=\left(\frac{1+\lambda z_2}{(1+\lambda z_2)^{3/2}+\lambda \sqrt
{z_1}}\right)^2\\[3mm]
    z_2=\left(\frac{1+\lambda z_1}{(1+\lambda
z_1)^{3/2}+\lambda \sqrt {z_2}}\right)^2.\\
    \end{array}%
\right.
\eqno(13)$$

Введем обозначения $y=\sqrt{z_1}$ и $x=\sqrt {z_2}$. Тогда
систему уравнений (13) можно переписать следующим образом:
$$
\left\{%
\begin{array}{ll}
  y=\frac{1+\lambda x^2}{(1+\lambda x^2)^{3/2}+\lambda y}\\[3mm]
    x=\frac{1+\lambda y^2}{(1+\lambda y^2)^{3/2}+\lambda x}.\\
    \end{array}%
\right.
\eqno(14)$$
Ясно, что $x<1$ и $y<1$.
Вычтев из первого уравнения (14) второе, после некоторых преобразований, получим
$$(y-x)(A+\lambda+\lambda^2(x^2+xy+y^2))=\lambda (y-x)(y+x)(1+\lambda
x^2)(1+\lambda y^2)\frac{1}{\sqrt{1+\lambda x^2}+\sqrt{1+\lambda
y^2}},$$
где
$$A=(\lambda x+(1+\lambda y^2)^{3/2})(\lambda y+(1+\lambda x^2)^{3/2}),$$
Отсюда $y=x$ или
$$
A+\lambda+\lambda^2(x^2+xy+y^2)=\lambda(y+x)(1+\lambda
x^2)(1+\lambda y^2)\frac{1}{\sqrt{1+\lambda x^2}+\sqrt{1+\lambda
y^2}}.
\eqno(15)$$
Аналогично предыдущему случаю при $y=x$ мы получим решение, которое соответствует единственной трансляционно-инвариантной мере Гиббса.

Пусть $y\neq x$. Тогда $x>y$ или $y>x$. Предположим, что $x>y$. Докажем, что уравнение (15) не имеет решений. Для этого достаточно показать справедливость следующего неравенства:
$$
(A+\lambda+\lambda^2(x^2+xy+y^2))(\sqrt{1+\lambda x^2}+\sqrt{1+\lambda y^2})> \lambda(y+x)(1+\lambda x^2)(1+\lambda y^2).
\eqno(16)$$
Для $A$ и $\sqrt{1+\lambda x^2}+\sqrt{1+\lambda y^2}$ воспользуемся неравенствами Бернулли и Коши:
$$A=(\lambda x+(1+\lambda y^2)^{3/2})(\lambda y+(1+\lambda x^2)^{3/2})\geq
\left(1+\frac{3}{2}\lambda x^2+\lambda y\right)\left(1+\frac{3}{2}\lambda y^2+\lambda x\right),$$
$$\sqrt{1+\lambda x^2}+\sqrt{1+\lambda y^2}\geq
1+\frac {1}{2}\lambda x^2+1+\frac {1}{2}\lambda y^2\geq 2+\lambda xy.$$
Используя последние два неравенства, из (16) получим
$$\left(\left(1+\frac{3}{2}\lambda x^2+\lambda y\right)\left(1+\frac{3}{2}\lambda
y^2+\lambda x\right)+\lambda+\lambda^2(x^2+xy+y^2)\right)(2+\lambda
xy)>\lambda(y+x)(1+\lambda x^2)(1+\lambda y^2)$$
или
$$2+\lambda\left(3y^2+3x^2+x+y+xy+2\right)+$$
$$\lambda^2\left(\frac92 x^2y^2+2x^3+2y^3+4xy+2x^2+2y^2+xy\left(\frac32 y^2+\frac32 x^2+1\right)\right)+$$
$$+\lambda^3\left(\frac94 x^3y^3+\frac32 x^4y+\frac32 xy^4+2x^2(1-y^2)+2y^2+x^3y+xy^3\right)>0.$$
Здесь в выражении при $\lambda^3$ воспользовались неравенством $x^3y^2+x^2y^3<2x^2$, а это неравенство верно, т.к. $0<x<1$ и $0<y<1$. Отсюда получим справедливость неравенства (16). Следовательно, уравнение (15) не имеет решений. Случай $x<y$ доказывается аналогично.

\textbf{Случай $I_4$: $k=4$ и $i=1$.} В этом случае система уравнений (4) имеет вид ($z_1=y$, $z_2=x$):
$$
\left\{%
\begin{array}{ll}
  y=\frac{1+\lambda x}{(1+\lambda
x)^4+\lambda}\\[3mm]
    x=\frac{1+\lambda y}{(1+\lambda y)^4+\lambda}.\\
    \end{array}%
\right.
\eqno(17)$$
Ясно, что $x<1$ и $y<1$.
Вычтев из первого уравнения (17) второе, после некоторых преобразований, получим
$$(y-x)[(\lambda+(1+\lambda y)^4)(\lambda+(1+\lambda
x)^4)+\lambda^2]=\lambda(y-x)(1+\lambda x)(1+\lambda y)[(1+\lambda
x)^2+(1+\lambda x)(1+\lambda y)+(1+\lambda y)^2].$$
Отсюда $y=x$ или
$$
(\lambda+(1+\lambda y)^4)(\lambda+(1+\lambda
x)^4)+\lambda^2=\lambda(1+\lambda x)(1+\lambda y)[(1+\lambda
x)^2+(1+\lambda x)(1+\lambda y)+(1+\lambda y)^2].
\eqno(18)$$

При $y=x$ мы получим решение, которое соответствует единственной трансляционно-инвариантной мере Гиббса.
Докажем, что уравнение (18) не имеет решений. Для этого перепишем уравнение (18):
$$(1+\lambda x)^4(1+\lambda y)^4+\lambda (1+\lambda x)^4+\lambda (1+\lambda
y)^4+2\lambda^2=\lambda(1+\lambda x)(1+\lambda y)[(1+\lambda
x)^2+(1+\lambda x)(1+\lambda y)+(1+\lambda y)^2].$$
Пусть $y\neq x$. Тогда $x>y$ или $y>x$. Предположим, что $x>y$. Тогда $1+\lambda x>1+\lambda y$ $(\lambda>0)$ .
Чтобы доказать, что уравнение (18) не имеет решений, достаточно показать справедливость неравенства
$$(1+\lambda x)^4(1+\lambda y)^4+\lambda (1+\lambda x)^4+\lambda
(1+\lambda y)^4+2\lambda^2>\lambda(1+\lambda x)(1+\lambda
y)[(1+\lambda x)^2+(1+\lambda x)(1+\lambda y)+(1+\lambda y)^2].$$
Следующие неравенства очевидны:
$$\lambda (1+\lambda x)^4>\lambda(1+\lambda x)^3(1+\lambda y),$$
$$\lambda^2+(1+\lambda x)^4(1+\lambda y)^4\geq 2\lambda (1+\lambda
x)^2(1+\lambda y)^2>\lambda (1+\lambda x)^2(1+\lambda y)^2+\lambda
(1+\lambda x)(1+\lambda y)^3.$$
Следовательно,
$$(1+\lambda x)^4(1+\lambda y)^4+\lambda (1+\lambda x)^4+\lambda (1+\lambda
y)^4+2\lambda^2>\lambda(1+\lambda x)(1+\lambda y)[(1+\lambda
x)^2+(1+\lambda x)(1+\lambda y)+(1+\lambda y)^2],$$
что и требовалось доказать. Случай $x<y$ доказывается аналогично.

\textbf{Случай $I_4$: $k=4$ и $i=2$.} Система уравнений (4) на инвариантном множестве $I_4$ при $k=4$ и $i=2$
имеет вид ($y=\sqrt{z_1}$, $x=\sqrt {z_2}$):
$$
\left\{%
\begin{array}{ll}
  y=\frac{1+\lambda x^2}{(1+\lambda x^2)^2+\lambda y}\\[3mm]
    x=\frac{1+\lambda y^2}{(1+\lambda y^2)^2+\lambda x}.\\
    \end{array}%
\right.
\eqno(19)$$
Ясно, что $x<1$ и $y<1$ .
Аналогично предыдущему случаю, вычтев из первого уравнения (19) второе, после некоторых преобразований, получим
$$(y-x)[(\lambda y+1+2\lambda x^2+\lambda^2x^4)(\lambda x+1+2\lambda
y^2+\lambda^2y^4)+\lambda+\lambda^2(x^2+xy+y^2)]=$$
$$\lambda(y-x)(y+x)(1+\lambda(x^2+y^2)+\lambda^2x^2y^2).$$
Отсюда получим $y=x$ или уравнение
$$\lambda^4x^4y^4+\lambda^3(2x^2y^4+y^5+x^5+2x^4y^2)+\lambda^2(y^4+4x^2y^2+2x^3+x^4+2y^3+x^2+y^2+2xy)+$$
$$
\lambda(2x^2+2y^2+x+y+1)+1=\lambda^3(x^3y^2+x^2y^3)+\lambda^2(x^3+x^2y+xy^2+y^3)+\lambda(x+y).
\eqno(20)$$

При $y=x$ получим единственное решение из $I_1$.

Пусть $y\neq x$. Тогда $x>y$ или $y>x$. Предположим, что $x>y$. Аналогично случаю $k=3, i=2$, оценив выражения при $\lambda$, $\lambda^2$ и $\lambda^3$, можно показать, что левая часть равенства (20) строго больше правой части этого равенства, т.е. уравнение (20) не имеет решений при $x<1$ и $y<1$. А это значит, что в рассматриваемом случае слабо периодическая мера Гиббса единственна и она совпадает с единственной трансляционно-инвариантной мерой Гиббса. Случай $x<y$ доказывается аналогично.

\textbf{Случай $I_4$: $k=4$ и $i=3$.} В этом случае система
уравнений (4) на $I_4$ имеет вид:
$$
\left\{%
\begin{array}{ll}
    z_1=\left(\frac{1+\lambda z_2}{(1+\lambda z_2)^{4/3}+\lambda
z^{2/3}_1}\right)^3, \\[3mm]
    z_2=\left(\frac{1+\lambda z_1}{(1+\lambda z_1)^{4/3}+\lambda
z^{2/3}_2}\right)^3. \\
    \end{array}%
\right.
\eqno(21)$$
Ясно, что $0<z_1<1$ и $0<z_7<1$. Введем обозначения: $z_1=y^3$,
$z_2=x^3$. Тогда система уравнений (21) имеет вид:
$$
\left\{%
\begin{array}{ll}
   y=\frac{1+\lambda x^3}{(1+\lambda x^3)^{4/3}+\lambda y^2}, \\[3mm]
   x=\frac{1+\lambda y^3}{(1+\lambda y^3)^{4/3}+\lambda x^2}. \\
    \end{array}%
\right.
\eqno(22)$$
Ясно, что $x<1$ и $y<1$. Вычтев из первого уравнения (22)
второе, получим
$$y-x=\frac{(1+\lambda x^3)((1+\lambda y^3)^{4/3}+\lambda x^2)-(1+\lambda y^3)((1+\lambda x^3)^{4/3}+\lambda y^2)}{((1+\lambda y^3)^{4/3}+\lambda x^2)((1+\lambda
x^3)^{4/3}+\lambda y^2)}.$$
Введя обозначение
$$A=((1+\lambda y^3)^{4/3}+\lambda x^2)((1+\lambda
x^3)^{4/3}+\lambda y^2),$$
последнее уравнение можно переписать следующим образом:
$$(y-x)(A+\lambda(x+y)+\lambda^2(x^4+x^3y+x^2y^2+xy^3+y^4))=\frac{\lambda (y-x)(x^2+y^2+xy)(1+\lambda
x^3)(1+\lambda y^3)}{B}.$$
Здесь
$$B=(1+\lambda
x^3)^{2/3}+(1+\lambda x^3)^{1/3}(1+\lambda y^3)^{1/3}+(1+\lambda
y^3)^{2/3}.$$
Отсюда $y=x$ или
$$
B(A+\lambda(x+y)+\lambda^2(x^4+x^3y+x^2y^2+xy^3+y^4))=\lambda(x^2+y^2+xy)(1+\lambda
x^3)(1+\lambda y^3).
\eqno(23)$$
Аналогично предыдущему случаю при $y=x$ мы
получим решение, которое соответствует единственной
трансляционно-инвариантной мере Гиббса.

Пусть $y\neq x$. Тогда $x>y$ или $y>x$. Предположим, что $x>y$.
Докажем, что уравнение (23) не имеет решений. Для этого
достаточно показать, что левая часть уравнения (23) строго больше правой части, т.е.
$$
B(A+\lambda(x+y)+\lambda^2(x^4+x^3y+x^2y^2+xy^3+y^4))>\lambda(x^2+y^2+xy)(1+\lambda
x^3)(1+\lambda y^3).
\eqno(24)$$

Для $A$ и $B$ воспользуемся неравенствами Бернулли и Коши из курса математического анализа:
$$A=((1+\lambda y^3)^{4/3}+\lambda x^2)((1+\lambda
x^3)^{4/3}+\lambda y^2)\geq\left(1+\frac43\lambda y^3+\lambda
x^2\right)\left(1+\frac43\lambda x^3+\lambda y^2\right),$$
$$A> 1+\lambda(x^2+x^3+y^2+y^3)+\lambda^2(x^5+x^2y^2+x^3y^3+y^5),$$
$$B=(1+\lambda x^3)^{2/3}+(1+\lambda x^3)^{1/3}(1+\lambda y^3)^{1/3}+(1+\lambda
y^3)^{2/3}\geq$$
$$\left(1+\frac13\lambda x^3\right)^2+\left(1+\frac13\lambda
x^3\right)\left(1+\frac13\lambda y^3\right)+\left(1+\frac13\lambda y^3\right)^2,$$
$$B\geq3+\lambda(x^3+y^3)+\frac19\lambda^2(x^6+x^3y^3+y^6)>3+\lambda(x^3+y^3).$$
Используя эти неравенства, из (24) получим
$$3+\lambda(4x^3+4y^3+3x^2+3y^2+3x+3y)+$$
$$\lambda^2((x^3+y^3)(x^2+x^3+y^2+y^3+x+y)+3(x^5+2x^2y^2+x^3y^3+y^5+x^4+x^3y+xy^3+y^4))+$$
$$\lambda^3(x^3+y^3)(x^5+2x^2y^2+x^3y^3+y^5+x^4+x^3y+xy^3+y^4)>$$
$$>\lambda(x^2+xy+y^2)+\lambda^2(x^3+y^3)(x^2+xy+y^2)+\lambda^3x^3y^3(x^2+xy+y^2).$$
В обеих частях последнего неравенства, оценив выражения при $\lambda$, $\lambda^2$ и $\lambda^3$ при условиях $x>y$, $0<x<1$ и $0<y<1$ получим справедливость (24). При этом в выражениях при $\lambda$, $\lambda^2$ и $\lambda^3$ используются неравенства $x^2+xy+y^2<3x$, $x^2+xy+y^2<x^2+y^2+x$ и $x^5y^3+x^4y^4+x^3y^5<x^8+x^4y^3+x^3y^4$, соответственно. Значит, уравнение (23) не имеет решений. Следовательно, в рассматриваемом случае слабо периодическая мера Гиббса единственна и она совпадает с единственной трансляционно-инвариантной мерой Гиббса. Случай $x<y$ доказывается аналогично.

\textbf{Случай $I_4$: $k=5$ и $i=1$.} В этом случае система
уравнений (4) на $I_4$ имеет вид ($y=z_1$, $x=z_2$):
$$
\left\{%
\begin{array}{ll}
   y=\frac{1+\lambda x}{(1+\lambda x)^5+\lambda}, \\[3mm]
   x=\frac{1+\lambda y}{(1+\lambda y)^5+\lambda}. \\
    \end{array}%
\right.
\eqno(25)$$
Ясно, что $0<x<1$ и $0<y<1$. Вычтев из первого уравнения (25) второе, после некоторых преобразований, будем иметь
$$A(y-x)=(y-x)[\lambda (1+\lambda x)((1+\lambda y)\cdot[(1+\lambda y)+(1+\lambda x)]\cdot[(1+\lambda y)^2+(1+\lambda x)^2]-\lambda^2],$$
где
$$A=((1+\lambda x)^5+\lambda)((1+\lambda y)^5+\lambda).$$
Отсюда $x=y$ или
$$
A=\lambda (1+\lambda x)((1+\lambda y)\cdot[(1+\lambda y)+(1+\lambda x)]\cdot[(1+\lambda y)^2+(1+\lambda x)^2]-\lambda^2.
\eqno(26)$$
В случае $x=y$ мы получим решение, которое соответствует единственной трансляционно-инвариантной мере Гиббса.

Пусть $y\neq x$. Покажем, что уравнение (26) не имеет решений при любых $\lambda>0$, $0<x<1$ и $0<y<1$.
Введем обозначения: $s=1+\lambda x$ и $t=1+\lambda y$. Ясно, что $s>1$, $t>1$
и $A=(s^5+\lambda)(t^5+\lambda)$. Тогда уравнение (26) после некоторых преобразований будет иметь следующий вид:
$$s^5t^5+\lambda s^5+\lambda t^5+2\lambda^2=\lambda(s^4t+s^3t^2+s^2t^3+st^4).$$
Пользуясь неравенством Коши ($s^5t^5+2\lambda^2\geq2\sqrt2\lambda s^{5/2}t^{5/2}>2.5s^{5/2}t^{5/2}$), получим
$$s^5t^5+\lambda s^5+\lambda t^5+2\lambda^2>\lambda (s^5+2.5 s^{5/2}t^{5/2}+t^5)$$
Чтобы доказать, что уравнение (26) не имеет решений, достаточно показать справедливость неравенства
$$
s^5+2.5s^{5/2}t^{5/2}+t^5> s^4t+s^3t^2+s^2t^3+st^4.
\eqno(27)$$
Так как $s\neq t$, то предположим, что $s>t$. Обе стороны неравенства (27) разделим на $t^5$ и
обозначим: $\frac s t=q$. Очевидно, что $q>1$. Тогда
$$q^5+2.5 q^{5/2}+1>q^5+2.5 q^2+1>q^4+q^3+q^2+q$$
или
$$g(q)=q^5-q^4-q^3+1.5q^2-q+1>0.$$
Легко доказать, что функция $g(q)>0$ при любых $q>1$.
Действительно,производная этой функции при $q>1$
$$g'(q)=5q^4-4q^3-3q^2+3q-1=(q-1)(5q^3+q^2-2q+1)=(q-1)(3q^3+q^2+2q(q^2-1)+1)>0$$
и $g(1)=0.5>0$. Из всегосказанного следует, что уравнение (26) не имеет решений при любых $\lambda>0$, $0<x<1$ и $0<y<1$. Случай $s<t$ доказывается аналогично. Теорема доказана.

\section{Условия неединственности слабо периодических мер}\

\textbf{Случай $I_2$.} Пусть $k=3$ и $i=1$. Тогда на $I_2$ система уравнений (4)
имеет вид:
$$
\left\{%
\begin{array}{ll}
    z_1=\frac{(1+\lambda z_1)^3}{(1+\lambda
z_1)^3+\lambda \sqrt {z_2}} \cdot \frac {1}{(1+\lambda z_2)^2} \\[3mm]
    z_2=\frac{(1+\lambda z_2)^3}{(1+\lambda
z_2)^3+\lambda \sqrt {z_1}} \cdot \frac {1}{(1+\lambda z_1)^2}. \\
    \end{array}%
\right.
\eqno(28)$$

Введем обозначения: $x=1+\lambda z_1$, $y=1+\lambda z_2$.
Тогда (28) будет иметь вид
$$
\left\{%
\begin{array}{ll}
    y^2=\frac{\lambda x^3}{(x^3+\lambda)(x-1)} \\[3mm]
    x^2=\frac{\lambda y^3}{(y^3+\lambda)(y-1)}, \\
    \end{array}%
\right.
\eqno(29)$$
или
$$
\left\{%
\begin{array}{ll}
    x=h(y) \\
    y=h(x), \\
    \end{array}%
\right.
\eqno(30)$$
где
$$h(x)=\sqrt{\frac{\lambda x^3}{(x^3+\lambda)(x-1)}}.$$
Из $\lambda>0$, $z_1>0$ и $z_2>0$ следует, что  $x>1$ и
$y>1$.

Следующая лемма очевидна.

\textbf{Лемма 3.} Если  $(x_0, y_0)$ является решением системы уравнений (30),
то $(y_0, x_0)$ тоже является решением системы уравнений (30).

В частности, из леммы 3 следует, что если существует решение $(x_0, y_0)$ ($x_0\neq y_0$), то (30) имеет более одного решения.

\textbf{Следствие.} Если количество решений уравнения $x=h(x)$
является нечетным (четным), то количество решений (30) тоже
нечетное (четное).

\textbf{Замечание 3.} В работе [13] при доказательстве пункта 3 теоремы 1 рассматривалось уравнение
$$f(x,\lambda)=x^{16}-(\lambda+4)x^{15}+3(\lambda+2)x^{14}-4x^{13}+(1-14\lambda)x^{12}+3\lambda(\lambda+8)x^{11}-$$
$$16\lambda x^{10}-4\lambda(5\lambda-1)x^9+36\lambda^2x^8+\lambda^2(\lambda-24)x^7+\lambda^2(6-13\lambda)x^6+$$
$$24\lambda^3x^5-16\lambda^3x^4+\lambda^3(4-3\lambda)x^3+6\lambda^4x^2-4\lambda^4x+\lambda^4=0,$$
левая часть которого при значении $\lambda=1.8$ пересекала ось $Ox$ четыре раза и все значения $x$ были больше 1:
$x_1\approx1.516308807$, $x_2\approx1.285720838$, $x_3\approx1.846900632$ и $x_4\approx2.150852569$,
но одно из соответствующих значений $y$ будет меньше 1: $y_1\approx1.516308807$, $y_2\approx0.9872212333$, $y_3\approx1.183212942$, $y_4\approx1.234609281$. Кроме того, одно из этих решений $(x_1,y_1)$ соответствует трансляционно-инвариантной мере Гиббса. Поэтому утверждение 3-пункта теоремы 1 должно было быть так: при $\lambda>\lambda_{0}$ существуют не менее трех мер Гиббса, одна из которых является трансляционно-инвариантной, а остальные два слабо периодическими (не трансляционно-инвариантными) мерами Гиббса.

Далее, из первого уравнения (29) найдем $y$ и подставим во второе уравнение.
Тогда после некоторых преобразований получим уравнение
$$
\lambda x^6-((x^3+\lambda)(x-1))^2=x\sqrt{\lambda x(x^3+\lambda)(x-1)}(\lambda x+x^3-(x^3+\lambda)(x-1)).
\eqno(31)$$
Здесь необходимо, чтобы выражения в левой и правой части уравнения (31)
$$
\lambda x^6-((x^3+\lambda)(x-1))^2,
\eqno(32)$$
$$
\lambda x+x^3-(x^3+\lambda)(x-1)
\eqno(33)$$
имели одинаковые знаки. Перепишем выражение (32) следующим образом:
$$\lambda x^6-((x^3+\lambda)(x-1))^2=-\lambda^2(x-1)^2+\lambda
x^3(x^3-2(x-1)^2)-x^6(x-1)^2.$$
Тогда получим квадратный трехчлен относительно переменной $\lambda$, который имеет корни вида:
$$\acute{\lambda_1}(x)=\frac{x^3(x^3-2(x-1)^2)-x^3\sqrt{x^6-4x^3(x-1)^2}}{2(x-1)^2},$$
$$\acute{\lambda_2}(x)=\frac{x^3(x^3-2(x-1)^2)+x^3\sqrt{x^6-4x^3(x-1)^2}}{2(x-1)^2}.$$
Ясно, что при $\acute{\lambda_1}(x)<\lambda<\acute{\lambda_2}(x)$ выражение (32) будет положительным,
в противном случае отрицателным.

Далее, перепишем выражение (33)
$$\lambda x+x^3-(x^3+\lambda)(x-1)=\lambda+2x^3-x^4.$$
Отсюда, если $\lambda>\breve{\lambda}(x)=x^4-2x^3$, то (33) будет положительным, в противном случае
отрицательным. Значит, одновременно должны выполнятся условия:
$$
\acute{\lambda_1}(x)<\lambda<\acute{\lambda_2}(x), \ \lambda>\breve{\lambda}(x)
\eqno(34)$$
или
$$
\lambda<\acute{\lambda_1}(x), \ \lambda>\acute{\lambda_2}(x), \ \lambda<\breve{\lambda}(x).
\eqno(35)$$
Теперь, обе стороны уравнения (31) возведем в квадрат. Тогда получим уравнение
$$f(\lambda,x)=x^{16}-(\lambda+4)x^{15}+3(\lambda+2)x^{14}-4x^{13}+(1-14\lambda)x^{12}+3\lambda(\lambda+8)x^{11}-$$
$$-16\lambda x^{10}-4\lambda(5\lambda-1)x^9+36\lambda^2x^8+\lambda^2(\lambda-24)x^7+\lambda^2(6-13\lambda)x^6+$$
$$
+24\lambda^3x^5-16\lambda^3x^4+\lambda^3(4-3\lambda)x^3+6\lambda^4x^2-4\lambda^4x+\lambda^4=0.
\eqno(36)$$
Заметим, что если $y=x$, то решения системы уравнений (29) соответствуют решениям из инвариантного множества $I_1={(z_1,z_2,z_7,z_8)\in R^4:z_1=z_2=z_7=z_8}$, а на $I_1$ известно, что решение системы уравнений (4), и значит, решение системы уравнений (29) единственно. Кроме того, это решение находится среди решений (36).
При $y=x$ из (29) получим
$$\lambda_1(x)=\lambda=-x^3+x^4.$$
Рассмотрим уравнение (36) относительно переменной $\lambda$ и разделим его на многочлен $\lambda+x^3-x^4$. В результате будем иметь уравнение вида
$$g(\lambda,x)=(-3x^3+6x^2-4x+1)\lambda^3+(-2x^7-4x^6+14x^5-11x^4+3x^3)\lambda^2+$$
$$+(x^{11}-2x^{10}-2x^9+11x^8-10x^7+3x^6)\lambda-x^{12}+3x^{11}-3x^{10}+x^9=0.$$
Решив уравнение $g(\lambda,x)=0$ с помощью формулы Кардано, получим следующие решения:
$$\lambda_2(x)=\frac{x^3(x-1)^3}{3x^2-3x+1},$$
$$\lambda_3(x)=\frac{x^3(2-x-x^2+x \sqrt{x^2+2x-3})}{2x-2},$$
$$\lambda_4(x)=\frac{x^3(2-x-x^2-x \sqrt{x^2+2x-3})}{2x-2}$$
Ясно, что $\lambda_4(x)<0$, т.к. $x>1$.

Рассмотрим $\lambda=\lambda_2(x)$. Заметим, что $\lambda_2(x)$ при $x>1$ должно одновременно удовлетворяет условиям
(34) или (35). Проверим условия (34).

Пусть $\acute{\lambda_1}(x)<\lambda_2(x)$. Тогда из этого неравенства, после некоторых преобразований, получим
$$4x^3(x-1)^4(2x^3-6x^2+4x-1)>0.$$
Следовательно, при $x>1$ будем иметь неравенство $2x^3-6x^2+4x-1>0$.

Пусть, теперь, $\lambda_2(x)<\acute{\lambda_2}(x)$, т.е.
$$\frac{x^3(x-1)^3}{3x^2-3x+1}<\frac{x^3(x^3-2(x-1)^2)+x^3\sqrt{x^6-4x^3(x-1)^2}}{2(x-1)^2}.$$
Отсюда после некоторых преобразований получим
$$2(x-1)^5<(3x^2-3x+1)(x^3-2(x-1)^2+\sqrt{x^6-4x^3(x-1)^2}).$$
Чтобы доказать справедливость последнего неравенства достаточно показать, что верно
$$2(x-1)^5<(3x^2-3x+1)(x^3-2(x-1)^2.$$
Действительно, последнее неравенство эквивалентно неравенству
$$x^3(x^2+x-1)>0,$$
которое верно при любых значениях $x>1$. Значит, условие $\acute{\lambda_1}(x)<\lambda_2(x)<\acute{\lambda_2}(x)$ эквивалентно условию $2x^3-6x^2+4x-1>0$.

Далее, проверим условие $\lambda_2(x)>\breve{\lambda}(x)$, т.е.
$$\frac{x^3(x-1)^3}{3x^2-3x+1}>x^4-2x^3.$$
Это неравенство эквивалентно неравенству
$2x^3-6x^2+4x-1<0$. Значит, $\lambda_2(x)$ одновременно не удовлетворяет условиям (34).
Заметим, что условия (35) одновременно не выполняются, т.к. $\lambda_2(x)<\acute{\lambda_2}(x)$ при любых значениях $x>1$. Значит, если $\lambda=\lambda_2(x)$, то не существует ни одного значения $x$, которое соответствовало бы значению $\lambda_2(x)$.

Пусть теперь $\lambda=\lambda_3(x)$. Проверим, удовлетворяет ли $\lambda_3(x)$ условиям (34):
$$\acute{\lambda_1}(x)<\lambda_3(x)<\acute{\lambda_2}(x), \ \lambda_3(x)>\breve{\lambda}(x).$$
Пусть $\acute{\lambda_1}(x)<\lambda_3(x)$. Тогда
$$\frac{x^3(x^3-2(x-1)^2)-x^3\sqrt{x^6-4x^3(x-1)^2}}{2(x-1)^2}<\frac{x^3(2-x-x^2+x
\sqrt{x^2+2x-3})}{2x-2}.$$
Отсюда после некоторых преобразований получим
$$4x^2(x-1)(2x^3-6x^2+4x-1)<0,$$
т.е. условие $\acute{\lambda_1}(x)<\lambda_3(x)$ эквивалентно условию $2x^3-6x^2+4x-1<0$.

Пусть теперь $\lambda_3(x)<\acute{\lambda_2}(x)$. Тогда имеем неравенство
$$\frac{x^3(2-x-x^2+x
\sqrt{x^2+2x-3})}{2x-2}<\frac{x^3(x^3-2(x-1)^2)+x^3\sqrt{x^6-4x^3(x-1)^2}}{2(x-1)^2},$$
которое эквивалентно неравенству
$$
2x^3-2x^2+x>x(x-1)\sqrt{x^2+2x-3}-\sqrt{x^6-4x^3(x-1)^2}.
\eqno(37)$$
Легко проверить, что правая часть неравенства (37) отрицательная при $1<x<\frac{3+\sqrt3}{2}$.

Пусть $x>\frac{3+\sqrt3}{2}$. Тогда из неравенства (37), после некоторых преобразований, получим
$$x^2+2>-(x+3)x^2(x^3-4x^2+8x-4).$$
Заметим, что это неравенство верно, т.к. $x^3-4x^2+8x-4>0$ при $x>1$.
Следовательно, условие $\acute{\lambda_1}(x)<\lambda_3(x)<\acute{\lambda_2}(x)$ выполняется при условии $2x^3-6x^2+4x-1<0$.

Далее, пусть $\lambda_3(x)>\breve{\lambda}(x)$, т.е.
$$\frac{x^3(2-x-x^2+x \sqrt{x^2+2x-3})}{2x-2}>x^4-2x^3,$$
а это неравенство эквивалентно неравенству
$$4(x-1)(2x^3-6x^2+4x-1)<0.$$
Отсюда неравенство $\lambda_3(x)>\breve{\lambda}(x)$ выполняется при условии $2x^3-6x^2+4x-1<0$.
Значит, условия (34) выполняются одновременно при всех значениях $x$, которые удовлетворяют неравенству $2x^3-6x^2+4x-1<0$.

Аналогичном образом можно показать, что условия (35) выполняются одновременно при всех значениях $x$, которые удовлетворяют неравенству $2x^3-6x^2+4x-1>0$, иными словами $\lambda_3(x)$ одновременно удовлетворяет условиям (34) или (35).

Найдем критические точки функции $\lambda_3(x)$:
$$\lambda'_3(x)=\frac{x^2(2x-3)\left[(-x^2-x+2+xt(x))t(x)r(x)+8x(x-1)\right]}{(2x-2)^2t(x)}=0.$$
Здесь $t(x)=\sqrt{x^2+2x-3}$ и $r(x)=\sqrt{2x^2+3x-3}+\sqrt{(2x+1)t(x)}$.
Легко показать, что
$$(-x^2-x+2+xt(x))t(x)r(x)+8x(x-1)>0.$$
Значит, функция $\lambda_3(x)$ имеет критические точки: $x=0$ и $x=3/2$. Отсюда, при $1<x<3/2$ функция $\lambda_3(x)$ убывает, а при $x>3/2$ она возрастает, т.е. $x=3/2$ есть точка минимума функции $\lambda_3(x)$ (см. Рис.1). Значит, минимальное значение этой функции есть:
$$\lambda_3(3/2)\equiv\lambda_{cr}=27/16.$$

\begin{center}
\includegraphics[width=6cm]{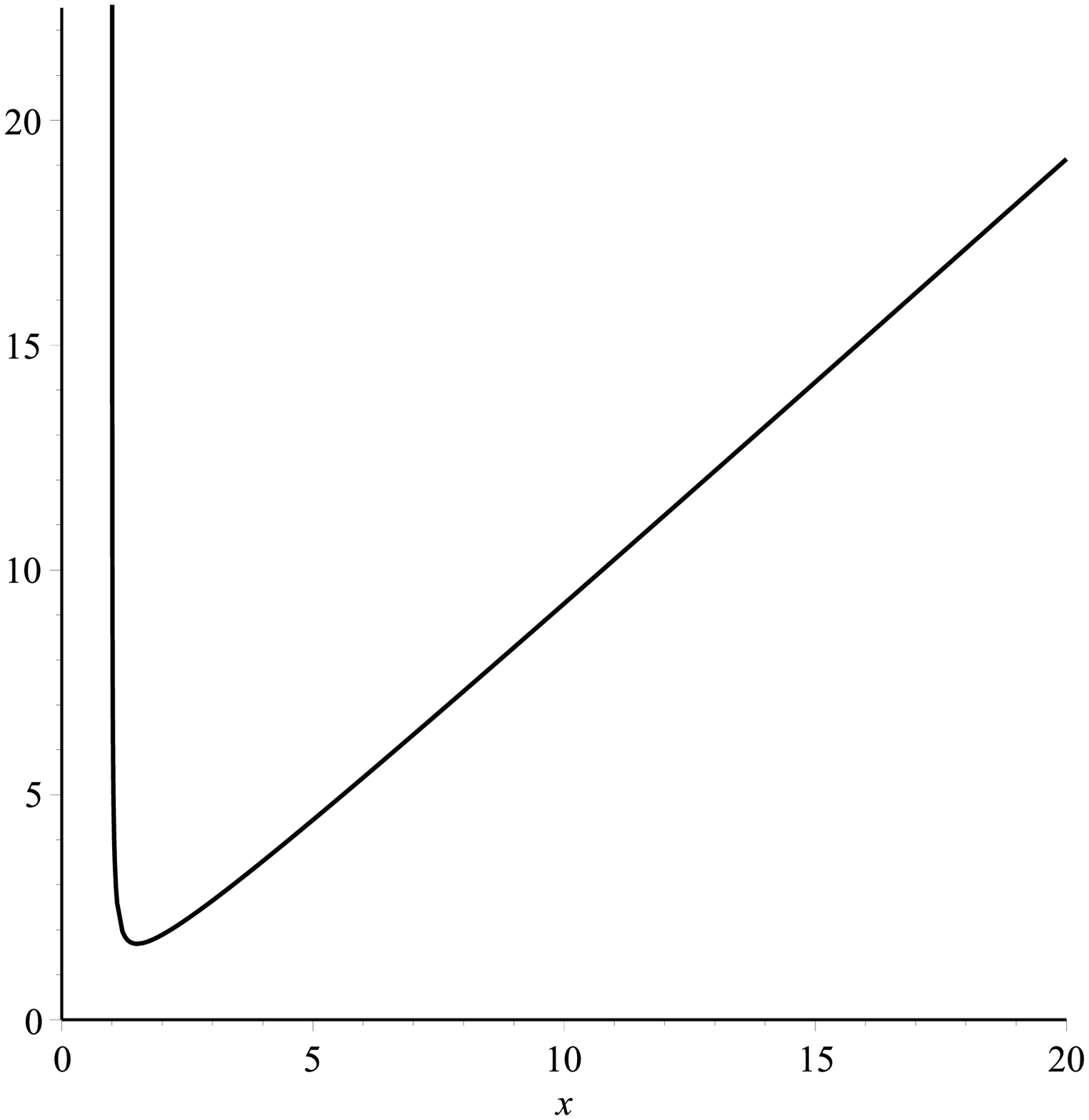}
\end{center}
\begin{center}{\footnotesize \noindent
 Рис. 1. График функции $\lambda_3(x)$.}\
\end{center}
Рассмотрим вторую производную функции $\lambda_3(x)$:
$$\lambda''_3(x)=\frac{s(x)}{(x-1)^2(x+3)\sqrt{(x-1)(x+3)}}.$$
Здесь
$$s(x)=6x^5+15x^4-12x^3-54x^2+18x+27-(4x^4+7x^3-21x^2-21x-9)\sqrt{(x-1)(x+3)}.$$
Покажем, что $\lambda''_3(x)>0$. Для этого достаточно показать справедливость следующего неравенства:
$$6x^5+15x^4-12x^3-54x^2+18x+27>(4x^4+7x^3-21x^2-21x-9)\sqrt{(x-1)(x+3)}.$$
Для этого увеличим правую часть этого неравенства следующим образом:
$$(4x^4+7x^3-21x^2-21x-9)\sqrt{(x-1)(x+3)}<(4x^4+7x^3-21x^2-21x-9)(x+3).$$
В результате получим неравенство
$$2(x+1)(x(x-1)^3-6x^2+19x+27)>0.$$
Легко показать, что $x(x-1)^3-6x^2+19x+27>0$ при $x>1$. Значит, $\lambda''_3(x)>0$.

Из всего сказанного следует, что каждому значению $\lambda=\lambda_3(x)$ соответствуют ровно два значения $x_1, x_2$ при $\lambda>(\lambda_3(x))_{min}=\lambda_{cr}$, одно значение $x_0$ при $\lambda=\lambda_{cr}$ и ни одного значения при $0<\lambda<\lambda_{cr}$.

С другой стороны, если в (29) из второго уравнения найдем $y$ и подставим в первое уравнение, то относительно $y$ получим точно такое же уравнение как уравнение (31). Подобно уравнению (31), анализируя это уравнение, можно получить, что оно будет иметь решение $y_0$ при $\lambda\leq\lambda_{cr}$ и решения вида $y_0, y_1, y_2$ при $\lambda>\lambda_{cr}$. Легко заметить, что $x_0=y_0, x_1=y_1$, $x_2=y_2$ и $\lambda_1(3/2)=\lambda_{cr}$.

Если $x_i=y_i, i=1,2$, то из из первого уравнения (30) получим, что $x_i=h(y_i)=h(x_i)$, т.е. $x_i=y_i=x_0, i=1,2$, т.к. уравнение $x_i=h(x_i)$ имеет единственное решение $x_0$ при любых значениях $\lambda>0$. Иными словами, в этих случаях мы получим для НС модели единственную трансляционно-инвариантную меру Гиббса, соответствующую решению $(x_0, x_0)$.

Если $x_i\neq y_i, i=1,2$, то в силу симметрии системы уравнений (29) и согласно лемме 3 получим ровно два решения $(x_2,y_1)$ и $(y_1,x_2)$, которые соответствуют слабо периодическим (не периодическим) мерам Гиббса для НС модели.

Итак, система уравнений (29) имеет единственное решение $(x_0,x_0)$ при $0<\lambda\leq\lambda_{cr}$, а при $\lambda>\lambda_{cr}$  имеет ровно три решения $(x_0,x_0)$, $(x_2,y_1)$, $(y_1,x_2)$.

Таким образом, верна следующая теорема.

\textbf{Теорема 4.} \textit{Пусть $k=3, i=1, \lambda_{cr}={27\over16}$. Тогда для HC-модели в случае нормального
делителя индекса четыре при $\lambda\leq\lambda_{cr}$ существует одна слабо
периодическая мера Гиббса (соответствующая совокупности величин из $I_2$), которая является
трансляционно-инвариантной, а при $\lambda>\lambda_{cr}$ существуют ровно три слабо периодические меры Гибсса (соответствующие совокупности величин из $I_2$), одна из которых является трансляционно-инвариантной, а две другие слабо периодическими
(не периодическими).}

\textbf{Случай $I_4$.} В этом случае полезна следующая лемма.

\textbf{Лемма 4.} (Кестен) [23] \textit{Пусть $f:[0,1]\rightarrow
[0,1]-$ непрерывная функция с неподвижной точкой $\xi \in (0,1)$.
Допустим, что $f$ дифференцируема в точке $\xi$ и $f^{'}(\xi)<-1.$
Тогда существуют точки $x_1, \ x_2, \ 0\leq x_1<\xi<x_2 \leq1$
такие, что $f(x_1)=x_2$ и $f(x_2)=x_1.$}

Пусть
$$s^{\pm}:=s^{\pm}(k)=\frac{k-3\pm\sqrt{k^2-6k+1}}{4},$$
$$\lambda^{\pm}:=\lambda^{\pm}(k)=(s^{\pm}+1)^ks^{\pm}.$$

Справедлива следующая теорема.

\textbf{Теорема 5.} При $k\geq6$, $i=1$ и $\lambda\in(\lambda^{-}(k), \lambda^{+}(k))$ для HC-модели в случае нормального делителя индекса четыре существуют не менее трех слабо периодических мер Гиббса, соответствующих совокупности величин из $I_4$. При этом одна из них является трансляционно-инвариантной, другие слабо периодическими (не периодическими) мерами Гиббса.

\textbf{Доказательство.} При $k\geq6$ и $i=1$ на $I_4$ система уравнений (4)
имеет вид ($x=z_1$ и $y=z_2$):
$$
\left\{%
\begin{array}{ll}
    x=\gamma(y) \\
    y=\gamma(x), \\
    \end{array}%
\right.
\eqno(38)$$
где
$$\gamma(x)=\frac{1+\lambda x}{(1+\lambda x)^k+\lambda}.$$
Ясно, что $0<x<1$ и $0<y<1$. Отсюда $0<\gamma(x)<1$, т.е. $\gamma:[0,1]\rightarrow[0,1]$.
Кроме того, $\gamma(x)-$непрерывная, дифференцируемая функция на отрезке $[0,1]$.
Известно, что уравнение $\gamma(x)=x$ имеет единственную неподвижную точку $x=\xi$.
Тогда
$$\xi=\frac{1+\lambda \xi}{(1+\lambda \xi)^k+\lambda}.$$
Отсюда после некоторых преобразований получим
$$
\xi=\frac{1}{(1+\lambda \xi)^k}.
\eqno(39)$$
Рассмотрим производную $\gamma'(\xi)$:
$$\gamma'(\xi)=\frac{\lambda(1+\lambda \xi)^k+\lambda^2-k\lambda(1+\lambda \xi)^k}{((1+\lambda
\xi)^k+\lambda)^2}.$$
Используя (39), получим
$$\gamma'(\xi)=\frac{\lambda \xi(1-k)+\lambda^2 \xi^2}{(1+\lambda \xi)^2}.$$
Тогда неравенство $\gamma'(\xi)<-1$ имеет вид
$$\lambda \xi(1-k)+\lambda^2 \xi^2+(1+\lambda \xi)^2<0.$$
Анализируя это неравенство можно увидеть, что оно имеет решение $\xi_1<\xi<\xi_2$, если условия теоремы выполнены, где
$$
\xi_1=\frac{k-3-\sqrt{k^2-6k+1}}{4\lambda}, \ \xi_2=\frac{k-3+\sqrt{k^2-6k+1}}{4\lambda}.
\eqno(40)$$
Тогда $\xi\in\left(\frac{s^-}\lambda,\frac{s^+}\lambda\right)$. Из (39) получим
$$
\lambda=\kappa(\xi):=\frac1\xi\left(\sqrt[k]{\frac1\xi}-1\right).
\eqno(41)$$
При $0<\xi<1$ имеем
$$\kappa'(\xi)=-\frac{k+1}{k\sqrt[k]{\xi^{2k+1}}}+\frac1{\xi^2}<0,$$
т.е. функция $\kappa(\xi)$ убывает. Кроме того, $\kappa(1)=0$, $\kappa(\xi)\rightarrow\infty$ при $\xi\rightarrow0$ и $\kappa'(\xi)=0$ при $\xi=\left(1+\frac{1}{k}\right)^k>1$. Отсюда при $\xi\in\left(\frac{s^-}\lambda,\frac{s^+}\lambda\right)$, в силу (40) и (41), имеем
$$\lambda\in(\kappa(\xi_2),\kappa(\xi_1))=(\lambda^{-}(k), \lambda^{+}(k)).$$
Следовательно, по лемме Кестена и лемме 3 получим, что система уравнений (38) при $\lambda^{-}(k)<\lambda<\lambda^{+}(k)$  имеет три решения $(\xi,\xi), \ (x_1,y_1)$ и $(y_1,x_1)$. Известно, что решение $(\xi,\xi)$ соответствует трансляционно-инвариантной мере Гиббса для НС-модели. Кроме того, такая мера единственна. Значит, решения $(x_1,y_1)$ и $(y_1,x_1)$ соответствуют слабо периодическим (не периодическим) мерам Гиббса для рассматриваемой модели. Теорема доказана.

\end{document}